\documentclass[journal]{IEEEtran}
\usepackage{graphicx}
\usepackage{subcaption}
\usepackage[usenames, dvipsnames]{color}
\usepackage{bm}
\usepackage{soul}

\ifCLASSINFOpdf
\else
\fi
\usepackage{algorithmic}
\usepackage{amsthm}

\newcommand{\cmmnt}[1]{}
\usepackage{array}

\usepackage{tabularx,booktabs}

\usepackage{stfloats}
\usepackage{url}
\usepackage{amsmath}
\usepackage{amsfonts}
\usepackage[numbers,sort&compress]{natbib}
\usepackage{bm}
\usepackage{hyperref}
\DeclareMathOperator{\argmax}{\arg\max}

\DeclareMathOperator{\hev}{H}

\begin{document}
\title{Towards a Theory of Systems Engineering Processes: A Principal-Agent Model of a One-Shot, Shallow Process}

\author{\IEEEauthorblockN{Salar Safarkhani\IEEEauthorrefmark{2},
Ilias Bilionis\IEEEauthorrefmark{2}\IEEEauthorrefmark{1}, Jitesh H. Panchal\IEEEauthorrefmark{2}}\\
\IEEEauthorblockA{\IEEEauthorrefmark{2}School of Mechanical Engineering, Purdue University, West Lafayette, Indiana 47907-2088}\\
\IEEEauthorblockA{\IEEEauthorrefmark{1}Corresponding author, email: ibilion@purdue.edu}
}
% <-this % stops a space
%\thanks{I. Bilionis is with the Department
%of Electrical and Computer Engineering, Georgia Institute of Technology, %Atlanta,
%\thanks{J. Doe and J. Doe are with Anonymous University.}% <-this % stops a #space
%\thanks{Manuscript received April 19, 2005; revised September 17, 2014.}}


\markboth{SUBMITTED TO IEEE SYSTEMS JOURNAL}%
{}



\maketitle

\begin{abstract}
Systems engineering processes coordinate the effort of different individuals to generate a product satisfying certain requirements. As the involved engineers are self-interested agents, the goals at different levels of the systems engineering hierarchy may deviate from the system-level goals which may cause budget and schedule overruns. Therefore, there is a need of a systems engineering theory that accounts for the human behavior in systems design. 
Undertaking such an ambitious endeavor is clearly beyond the scope of any single paper due to the inherent difficulty of rigorously formulating the problem in its full complexity along with the lack of empirical data.
However, as experience in the physical sciences shows, a lot of knowledge can be generated by studying simple hypothetical scenarios which nevertheless retain some aspects of the original problem.
To this end, the objective of this paper is to study the simplest conceivable systems engineering process, a principal-agent model of a one-shot (single iteration), shallow (one level of hierarchy) systems engineering process. We assume that the systems engineer maximizes the expected utility of the system, while the subsystem engineers seek to maximize their expected utilities. Furthermore, the systems engineer is unable to monitor the effort of the subsystem engineer and may not have complete information about their types or the complexity of the design task. However, the systems engineer can incentivize the subsystem engineers by proposing specific contracts. To obtain an optimal incentive, we pose and solve numerically a bi-level optimization problem. Through extensive simulations, we study the optimal incentives arising from different system-level value functions under various combinations of effort costs, problem-solving skills, and task complexities. Our numerical examples show that, the passed-down requirements to the agents increase as the task complexity and uncertainty grow and they decrease with increasing the agents' costs.
\end{abstract}


% \fi
\IEEEpeerreviewmaketitle

\section{Introduction}

\IEEEPARstart{C}{ost} and schedule overruns plague the majority of large systems engineering projects across multiple industry sectors including \cmmnt{transportation \cite{trans_cost},}power~\cite{pow_cost}, defense~\cite{def_cost}, and space~\cite{nasa_cost}.
%The systems engineering community has mostly attributed cost overruns to technical difficulties internal to systems engineering processes. 
%These include, for example, the adherence to requirement-based engineering as opposed to value-driven design~(ref) and the lack of internal incentive alignment~(ref).
%\IEEEPARstart{T}{he} systems engineering process (SEP) involves the coordination of several agents, e.g., managers, systems engineers (SE), subsystem engineers (sSE), contractors, and operators, to design, build, maintain, and retire complex engineering systems~\cite {incose, nasa}.
As design mistakes are more expensive to correct during the production and operation phases, the design phase of the systems engineering process (SEP) has the largest potential impact on cost and schedule overruns.
Collopy et al.~\cite{collopy} argued that requirements engineering (RE), which is a fundamental part of the design phase, is a major source of inefficiencies in systems engineering.
In response, they developed value-driven design (VDD)~\cite{collopy_vdd}, a systems design approach that starts with the identification of a system-level value function and guides the systems engineer (SE) to construct subsystem value functions that are aligned with the system goals.
According to VDD, the subsystem engineers (sSE) and contractors should maximize the objective functions passed down by the SE instead of trying to meet requirements.
\cmmnt{Since then, researchers have suggested various generalizations of VDD~\cite{Bloebaum,lee2014conceptual, kannan2015incorporation,kannan2017increased, bhatia2017integrating, salado2016integrating, bertoni2018evoke}, while applying it to many applications~\cite{papageorgiou2017value, vengadasalam2017value, shao2017bidirectional, cheung2012application, bertoni2017value}.}

RE and VDD make the assumption that the goals of the human agents involved in the SEP are aligned with the SE goals.
In particular, RE assumes that, agents attempt to maximize the probability of meeting the requirements, while VDD assumes that they will maximize the objective functions supplied by the SE.
However, this assumption ignores the possibility that the design agents, as all humans, may have personal agendas that not necessarily aligned with the system-level goals.
%In other words, the individuals in different levels of SEP hierarchy attempt to maximize their own expected utility rather than the system level expected utility. 

Contrary to RE and VDD, it is more plausible that the design agents seek to maximize their own objectives.
Indeed, there is experimental evidence that the quality of the outcome of a design task is strongly affected by the reward anticipated by the agent~\cite{reward, shergadwala2018quantifying, chaudhari2018analyzing}.
In other words, the agent decides how much effort and resources to devote to a design task after taking into account the potential reward.
In the field, the reward could be explicitly implemented as an annual performance-based bonus, or, as it is the case most often, it could be implicitly encoded in expectations about job security, promotion, professional reputation, etc. 
To capture the human aspect in SEPs, one possible way is one needs to follow a game-theoretic approach~\cite{malak}, \cite{cesun}.
Most generally, the SEP should can be modeled as a dynamical hierarchical network game with incomplete information.
Each layer of the hierarchy represents interactions among the SE and some sSEs, or the sSEs and other engineers or contractors.
With the term ``principal,'' we refer to any individual delegating a task, while we reserve the term ``agent'' for the individual carrying out the task.
Note that an agent may simultaneously be the principal in a set of interactions down the network.
For example, the sSE is the agent when considering their interaction with the SE (the principal), but the principal when considering their interaction with a contractor (the agent).
At each time step, the principals pass down delegated tasks along with incentives, the agents choose the effort levels that maximize their expected utility, perform the task, and return the outcome to the principals.

The iterative and hierarchical nature of real SEPs makes them extremely difficult to model in their full generality.
Given that our aim is to develop a theory of SEPs, we start from the simplest possible version of a SEP which retains, nevertheless, some of the important elements of the real process.
Specifically, the objective of this paper is to develop and analyze a principal-agent model of a one-shot, shallow SEP.
The SEP is ``one-shot'' in the sense that decisions are made in one iteration and they are final.
The term ``shallow'' refers to a one-layer-deep SEP hierarchy, i.e., only the SE (principal) and the sSEs (agents) are involved.
The agents maximize their expected utility given the incentives provided by the principal, and the principal selects the incentive structure that maximizes the expected utility of the system.
We pose this mechanism design problem~\cite{mechdesign} as a bi-level optimization problem and we solve it numerically.%~\cite{incentive, mechdesign}

A key component of our SEP model is the quality function of an agent.
The quality function is a stochastic process that models the principal's beliefs about the outcome of the delegated design task given that the agent devotes a certain amount of effort.
The quality function is affected by what the principal believes about the task complexity and the problem solving skills of the agent.
Following our work~\cite{asme}, we model the design task as a maximization problem where the agent seeks the optimal solution.
The principal expresses their prior beliefs about the task complexity by modeling the objective function as a random draw from a Gaussian process prior with a suitably selected covariance function.

As we showed in~\cite{asme}, conditioned on knowing the task complexity and the agent type, the quality function is well approximated by an increasing, concave function of effort with additive Gaussian noise. However, we will use a linear approximation for the quality function.

We study numerically two different scenarios.
The first scenario assumes that the SE knows the agent types and the task complexity, but they do not observe the agent's effort.
This situation is known in game theory as a \emph{moral hazard} problem~\cite{mirrlees}. 
The most common way to solve a moral hazard problem is to use the first order approach (FOA)~\cite{foa}. In the FOA, the incentive compatibility constraint of the agent is replaced by its first order necessary condition. 
However, the FOA depends on the convexity of the distribution function in effort which is not valid in our case. 
There have been several attempts to solve the principal-agent model where the requirements of the FOA may fail, nonetheless they must still satisfy the monotone likelihood ratio property~\cite{booth}.

In the second scenario, we study the case of moral hazard with simultaneous \emph{adverse selection}~\cite{myerson}, i.e., the SE observes neither the effort nor the type of agents nor the task complexity.
This is a Bayesian game with incomplete information.\cmmnt{~\cite{harsanyi}}. 
%The inability of the SE to get the sSEs to reveal their private, and possibly unfavorable, information about their problem-solving skills is known as adverse selection~\cite{myerson}.
%The SE's inability to monitor the effort of the sSEs causes an additional loss in their maximum expected utility.
%This is since the SE cannot directly choose the action that she would like the sSEs to perform~\cite{cvitanic}.
%Instead, the SE tries to reach her favorite action by offering a certain contract in which she pays for incentivizing the sSEs.
In this case,  the SE experiences additional loss in their expected utility, because the sSEs' can pretend to have different types.
%In this case, the offered contract must incentivize the sSEs to reveal their correct ability or problem complexity which denotes that the SE needs to pay more.
The revelation principle~\cite{revelation} guarantees that it suffices to search for the optimal mechanism within the set of incentive compatible mechanisms, i.e., within the set of mechanisms in which the sSEs are telling the truth about their types and technology maturity. In this paper, we solve the optimization problem in the principal-agent model, numerically with making no assumptions about the quality function.

The paper is organized as follows. In section \ref{model} we will derive the mathematical model of the SEP and we will study the type-independent and type-dependent optimal contracts. We will also introduce the value and utility functions. In section \ref{res}, we perform an exhaustive numerical study and show the solutions for several case studies. Finally, we conclude in section \ref{sec:concl}.

\section{Modeling a one-shot, shallow systems engineering process}
\label{model}

\subsection{Basic definitions and notation}
\label{moral}

As mentioned in the introduction, we develop a model of a one-shot (the game evolves in one iteration and the decisions are final), shallow (one-layer-deep hierarchy) SEP.
The SE has decomposed the system into $N$ subsystems and assigned a sSE to each one of them.
We use $i=1,\dots, N$ to label each subsystem.
From now on, we refer to the SE as the principal and the sSEs as the agents.
The principal delegates tasks to the agents along with incentives.
The agents choose how much effort to devote on their task by maximizing their expected utility.
The principal, anticipates this reaction and selects the incentives that maximize the system-level expected utility.
%Then the principal assigns each subsystem to an agent (sSE) and passes down the requirements alongside the incentive.
%Then, the agents choose their optimal level of efforts to maximize their own utility. 
%We will first, establish the optimal contract for subsystems in a full generality.
%To solve the optimization problem over the incentive function space we will expand the incentive function as linear combination of some basis functions. 

Let $\left(\Omega, \mathcal{F}, \mathbb{P}\right)$ be a probability space where, $\Omega$ is the sample space, $\mathcal{F}$ is a $\sigma$-algebra, and $\mathbb{P}$ is the probability measure.
With $\omega\in\Omega$ we refer to the random state of nature.
We use upper case letters for random variables (r.v.), bold upper case letters for their range, and lower case letters for their possible values.
For example, the type of agent $i$ is a r.v. $\Theta_i$ taking $M_i$ discrete values $\theta_i$ in the set $\boldsymbol{\Theta}_i \equiv \Theta_i(\Omega) = \{1,\dots,M_i\}$.
Collectively, we denote all types with the $N$-dimensional tuple $\Theta = (\Theta_1,\dots,\Theta_N)$ and we reserve $\Theta_{-i}$ to refer to the $(N-1)$-dimensional tuple containing all elements of $\Theta$ except $\Theta_i$.
This notation carries to any $N$-dimensional tuple.
For example, $\theta$ and $\theta_{-i}$ are the type values for all agents and all agents except $i$, respectively. The range of $\Theta$ is $\boldsymbol{\Theta} = \times_{i=1}^N\boldsymbol{\Theta}_i$.

The principal believes that the agents types vary independently, i.e., they assign a probability mass function (p.m.f.) on $\Theta$ that factorizes over types as follows:
\begin{equation}\label{eqn:principal_sok}
    \mathbb{P}[\Theta=\theta] = \prod_{i=1}^N\mathbb{P}[\Theta_i=\theta_i] = \prod_{i=1}^N p_{i\theta_i},
\end{equation}
for all $\theta$ in $\boldsymbol{\Theta}$, where $p_{ik}\ge 0$ is the probability that agent $i$ has type $k$,  for $k$ in $\boldsymbol{\Theta}_i$.
Of course, we must have $\sum_{k=1}^{M_j}p_{ik} = 1$, for all $i=1,\dots,N$.

Each agent knows their type, but their state of knowledge about all other agents is the same as the principal's.
That is, if agent $i$ is of type $\Theta_i=\theta_i$, then their state of knowledge about everyone else is captured by the p.m.f.:
\begin{equation}\label{eqn:agent_sok}
    \mathbb{P}[\Theta_{-i} = \theta_{-i} | \Theta_i=\theta_i] = \frac{\mathbb{P}[\Theta=\theta]}{\mathbb{P}[\Theta_i=\theta_i]} = \prod_{j\not=i}p_{j\theta_j}. 
\end{equation}

Agent $i$ chooses a normalized effort level $e_i \in [0,1]$ for his delegated task. 
We assume that this normalized effort is the percentage of an agent's maximum available effort. 
The units of the normalized effort depend on the nature of the agent's subsystem. 
If the principal and the agent are both part of same organization then the effort can be the time that the agent dedicates to the delegated task in a particular period of time, e.g., in a fiscal year.
On the other hand, if the agent is a contractor, then the effort can be the percentage of the available yearly budget that the contractor spends on the assigned task. 
We represent the monetary cost of the $i$-th agent's effort with the random process $C_i\left(e_i\right)$.
%This is the disutility inherent in human effort. 
In economic terms, $C_i\left(e_i\right)$ is the opportunity cost, i.e., the payoff of the best alternative project in agent could devote their effort.
In general, we know that the process $C_i(e_i)$ should be an increasing function of the effort $e_i$.
For simplicity, we assume that the cost of effort of the agents is quadratic,
\begin{equation}
C_i(e_i) := c_{i\Theta_i}e_i^2,
\label{eq:cost}
\end{equation}
with a type-dependent coefficient $c_{ik}>0$ for all $k$ in $\boldsymbol{\Theta}_i$.

The quality function of the $i$-th agent is a real valued random $Q_i(e_i) := Q_i(e_i)$ process paremeterized by the effort $e_i$.
The quality function models everybody's beliefs about the design capabilities of agent $i$.
The interpretation of the quality function is as follows.
If agent $i$ devotes to the task an effort of level $e_i$, then they produce a random outcome of quality $Q_i(e_i)$.
In our previous work \cite{asme}, we created a stochastic model for the quality function of a designer where we explicitly captured its dependence on the problem-solving skills of the designer and on the task complexity.
In that work, we showed that $Q_i(e_i)$ has increasing and concave sample paths, that its mean function is increasing concave, and the standard deviation is decreasing with effort, albeit mildly, it is independent of the problem-solving skills of the designer, and it only increases mildly with increasing task complexity. 
Examining the spectral decomposition of the process for various cases, we observed that it can be well-approximated by:
\begin{equation}
    Q_i(e_i) = q_{i\Theta_i}^0(e_i) + \sigma_{i\Theta_i}\Xi_i,
\end{equation}
where, for $k$ in $\boldsymbol{\Theta}_i$, $q_{ik}^0(e_i)$ is an increasing, concave, type-dependent mean quality function, $\sigma_{ik}>0$ is a type-dependent standard deviation parameter capturing the aleatory uncertainty of the design process, and $\Xi_i$ is a standard normal r.v.
If we further assume that the time window for design is relatively small, then the $q_{ik}^0(e_i)$ term can be approximated as a linear function. Therefore, we will assume that the quality function is:
\begin{equation}
     Q_i(e_i) = \kappa_{i\Theta_i} e_i + \sigma_{i\Theta_i}\Xi_i,
\label{eq:qual}
\end{equation}
where, $\kappa$ is inversely proportional to the complexity of the problem.
For instance, a large $\kappa$ corresponds to a low-complexity task while a $\kappa$ corresponds to a high-complexity task. The standard deviation parameter $\sigma$ captures the inherent uncertainty of the design process and depends on the maturity of the underlying technology.
In summary, an agent's type is characterized by the triplet cost-complexity-uncertainty.

From the perspective of the principal, the r.v.'s $\Xi_i$ are independent of the agents' types $\Theta_i$ as they represent the uncertain state of nature.
A stronger assumption that we employ is that the $\Xi_i$'s are also independent to each other.
This assumption is strong because it essentially means that the qualities of the various subsystems are decoupled.
Under these independence assumptions, the state of knowledge of the principal is captured by the following probability measure:
\begin{equation}\label{eqn:principal_complete_sok}
    \mathbb{P}\left[\Theta = \theta, \Xi \in \times_{i=1}^N\mathbf{B}_i\right] = \prod_{i=1}^N\left[p_{i\theta_i}\int_{\mathbf{B}_i}\phi(\xi_i)d\xi_i\right],
\end{equation}
for all $\theta\in\boldsymbol{\Theta}$ and all Borel-measurable $\mathbf{B}_i\subset \mathbb{R}$.
Assuming that all these are common knowledge, the state of knowledge of agent $i$ after they observe their type $\theta_i$ (but before they observe $\Xi_i$) is
\begin{equation}
    \begin{array}{ccc}
    \mathbb{P}\left[\Theta_{-i} = \theta_{-i}, \xi \in \times_{i=1}^N\mathbf{B}_i\middle| \Theta_i=\theta_i\right] &&\\
    = \frac{\mathbb{P}\left[\Theta = \theta, \xi \in \times_{i=1}^N\mathbf{B}_i\right]}{\mathbb{P}[\Theta_i=\theta_i]}&&\\
    =\mathbb{P}[\Theta_{-i}=\theta_{-i} |\Theta_i=\theta_i]\prod_{i=1}^N\left[\int_{\mathbf{B}_i}\phi(\xi_i)d\xi_i\right]&&
    \end{array}
\label{eq:prob}
\end{equation}

Finally, we use $\mathbb{E}[\cdot]$ to denote the expectation of any quantity over the state of knowledge of the principal as characterized by the probability measure of Eq.~(\ref{eqn:principal_complete_sok}).
That is, the expectation of any function $f(\Theta,\Xi)$ of the agent types $\Theta$ and the state of nature $\Xi$ is
\begin{equation}
    \mathbb{E}[f(\Theta,\Xi)] = \sum_{\theta\in\boldsymbol{\Theta}}\int_{\mathbb{R}^N}f(\theta,\xi)\prod_{i=1}^N\left[p_{i\theta_i}\phi(\xi_i)\right]d\xi.
\label{eq:exp}
\end{equation}
Similarly, we use the notation $\mathbb{E}_{i\theta_i}[\cdot]$ to denote the conditional expectation over the state of knowledge of an agent $i$ who knows that their type is $\Theta_i=\theta_i$.
This is the expectation $\mathbb{E}[\cdot|\Theta_i = \theta_i]$ with respect to the probability measure of Eq.~(\ref{eqn:agent_sok}) and we have:
\begin{equation}
    \mathbb{E}_{ik}[f(\Theta,\Xi)] = \sum_{\theta_{-i}\in\boldsymbol{\Theta}_{-i}}\int_{\mathbb{R}^N}f(\theta_i,\theta_{-i},\xi)\frac{\prod_{j=1}^N\left[p_{j\theta_j}\phi(\xi_j)\right]}{p_{i\theta_i}}d\xi.
\label{eq:exp_ik}
\end{equation}

%We derive the optimal contract (transfer function) for the certain SEP in two different scenarios. 
%First, we assume that the principal knows the agents' types and their utility. However, monitoring the effort of the agents is impossible for the principal and she needs to incentivize them with a proper contract to achieve her maximum utility. In economic terms, this scenario is known as \emph{moral hazard problem}. 
%Second, we assume that the types or/and utilities of the agents are hidden to the principal. In this situation, the quality function \emph{is not a common knowledge} and they are private information available only for the principal. In economic terms this problem is known as the \emph{adverse selection}.

\subsection{Type-independent optimal contracts}
\label{sec:single-contracts}
We start by considering the case where the principal offers a single take-it-or-leave-it contract independent of the agent type.
This is the situation usually encountered in contractual relationships between the SE and the sSEs within the same organization.
The principal offers the contract and the agent decides whether or not to accept it.
If the agent accepts, then they select their level of effort by maximizing their expected utility, they work on their design task, they return the outcome quality back to the principal, and they receive their reward. We show a schematic view of this type of contracts in Fig. \ref{t1}. 
%Each agent enters a contract with the principal.
A contract is a monetary \emph{transfer function} $t_i:\mathbb{R}\rightarrow\mathbb{R}$ that specifies the agent's \emph{compensation} $t_i(q_i)$ contingent on the quality level $q_i$.
Therefore, the payoff of the $i$-th agent is the random process:
\begin{equation}
\Pi_i(e_i) = t_i\left(Q_i\left(e_i\right)\right) - C_i(e_i).
\end{equation}
We assume that the agent knows their type, but they choose the optimal effort level ex-ante, i.e., they choose the effort level before seeing the state of the nature $\Xi_i$.
Denoting their monetary utility function by $U_i(\pi_i) = u_{i\Theta_i}(\pi_i)$, the $i$-th agent selects an effort level by solving:
\begin{equation}
\label{exsub}
{e_i}^*_{\theta_i} = \underset{e_i\in[0,1]}\argmax\ \mathbb{E}_{i\theta_i}\left[U_i\left(\Pi_i(e_i)\right)\right].
\end{equation}
%This is essentially a conditional expectation over the possible values of their quality function and it will be specified in Sec.~ X.

Let $Q_i^*$ be the r.v. representing the quality function that the principal should expect from agent $i$ if they act optimally, i.e.,
\begin{equation}
    Q_i^* = Q_i(e_{i_{\Theta_i}}^*).
\end{equation}
Then the system level value is a r.v. of the form
\begin{equation}
    V = v(Q^*),
\end{equation}
where $v : \mathbb{R}^N\rightarrow \mathbb{R}$ is a function of the subsystem outcomes $Q^*$.
We introduce the form of the value function, $v(q)$, in Sec. \ref{res}.
Note that, even though in this work the r.v. $V$ is assumed to be just a function of $Q^*$, in reality it may also depend on the random state of nature, e.g., future prices, demand for the system services.
Consideration of the latter is problem-dependent and beyond the scope of this work.
%In RE, $V$ yields zero if the requirements are not met and some positive payoff otherwise.
%In VDD, $V$ is the net present value of the future systems cash flows.

Given the system value $V$ and taking into account the transfers to the agents, the system-level payoff is the r.v.
\begin{equation}
\label{sepi}
\Pi_0 = V  -\sum_{i=1}^Nt_i(Q_i^*).
\end{equation}
If the monetary utility of the principal is $u_0(\pi_0)$, then they should select the transfer functions $t(\cdot) = (t_1(\cdot),\dots,t_N(\cdot))$ by solving:
\begin{equation}
\label{eq_seopt}
t^*\left(\cdot\right)=\underset{t\left(\cdot\right)}\argmax\ \mathbb{E}\left[u_0\left(\Pi_0\right)\right].
\end{equation}
However,  guarantee that they want to participate in the SEP, the expected utility of the sSEs must be greater than the expected utility they would enjoy if they participated in another project.
Therefore, the SE must solve Eq.~(\ref{eq_seopt}) subject to the \emph{participation constraints}:
\begin{equation}
\label{eq_partcons}
    \mathbb{E}_{i\theta_i}\left[U_i\left(\Pi_i\right)\right] \ge \bar{u}_{i\theta_i},
\end{equation}
for all possible values of $\theta_i$, and all $i=1,\dots,N$, where $\bar{u}_{i\theta_i}$ is known as the \emph{reservation utility} of agent $i$.

\subsection{Type-depdenent optimal contracts}
\label{sec:type_differentiation}
By offering a single transfer function, the principal is unable to differentiate between the various agent types when adverse selection is an issue.
That is, all agent types, independently of their cost, complexity, and uncertainty attributes, exactly the same transfer function.
In other words, with a single transfer function the principal is actually targeting the average agent.
This necessarily leads to inefficiencies stemming from problems such as paying an agent involved in a low-complexity task more than a same cost and uncertainty agent involved in a high-complexity task.

\begin{figure}
\centering
	\begin{subfigure}{0.48\textwidth}
		\includegraphics[width=\textwidth]{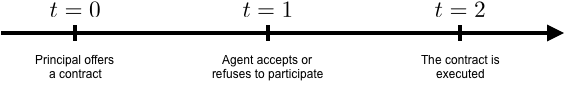}
        \caption{}
        \label{t1}
	\end{subfigure}
	\begin{subfigure}{0.48\textwidth}
		\includegraphics[width=\textwidth]{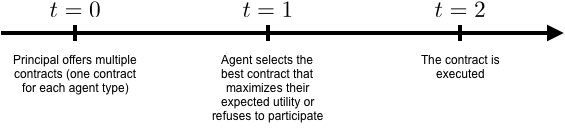}
        \caption{}
        \label{fig:timing2}
	\end{subfigure}
\caption{(a): Timing of the contract for type-independent contracts. (b): Timing of the contract for type-dependent contracts.}
\label{fig:timing}
\end{figure}

The principal can gain in efficiency by offering different transfer functions (if any exist) that target specific agent types.
For example, the principal could offer a transfer function that is suitable for cost-efficient, low-complexity, low-uncertainty agents, and one for cost-inefficient agents, low-complexity, low-uncertainty, etc., for any other combination that is supported by the principal's prior knowledge about the types of the agent population.
To implement this strategy the principal can employ the following extension to the mechanism of Sec.~\ref{sec:single-contracts}.
Prior to initiating work, the agents announce their types to the principal and they receive a contract that matches the announced type. In Fig. \ref{fig:timing2}, we show how this type of contract evolves in time.
Let us formulate this idea mathematically.
The $i$-th agent announces a type $\theta_i'$ in $\boldsymbol{\Theta}_i$ (not necessarily the same as their true type $\theta_i$), and they receive the associated, type-specific, transfer function $t_{i\theta_i'}(\cdot)$.
The payoff to agent $i$ is now:
\begin{equation}\label{agent_pay_as}
\Pi_i(e_i, \theta_i') = t_{i\theta_i'}(Q_i(e_i)) - C_i(e_i),
\end{equation}
where all other quantities are like before.
Given the announcement of a type $\theta_i'$, the rational thing to do for agent $i$ is to select a level of $e_i^*(\theta_i,\theta_i')$ by maximizing their expected utility, i.e., by solving:
\begin{equation}\label{agent_effort_atype_as}
e^*_{i\theta_i\theta_i'} = \underset{e_i\in[0,1]}{\arg\max}\mathbb{E}_{i\theta_i}[U_i(\Pi_i(e_i, \theta_i'))].
\end{equation}

Of course, the announcement of $\theta_i'$ is also a matter of choice and a rational agent should select also by maximizing their expected utility.
The obvious issue here is that agents can lie about their type.
For example, a cost-efficient agent (agent with low cost of effort) may pretend to be a cost-inefficient agent (agent with high cost of effort).
Fortunately, the revelation principle \cite{revelation} comes to the rescue and simplifies the situation.
It guarantees that, among the optimal mechanisms, there is one that is incentive compatible.
Thus it will be sufficient if the principal constraints their contracts to over truth-telling mechanisms.
Mathematically, to enforce truth-telling, the SE must satisfy the \emph{incentive compatibility} constraints:
\begin{equation}\label{ic}
\mathbb{E}_{i\theta_i}[U_i(\Pi_i(e_{i\Theta_i\theta_i}^*, \theta_i))]
\ge
\mathbb{E}_{i\theta_i}[U_i(\Pi_i(e_{i\Theta_i\theta_i'}^*, \theta_i'))],
\end{equation}
for all $\theta_i\not=\theta_i'$ in $\boldsymbol{\Theta}_i$.
Eq.~(\ref{ic}) expresses mathematically that ``the expected payoff of agent $i$ when they are telling the truth is always greater than or equal to the expected payoff they would enjoy if they lied.''

Similar to the developments of Sec.~\ref{sec:single-contracts}, the quality that the SE expects to receive is:
\begin{equation}
    Q_i^* = Q_i(e_{i\Theta_i\Theta_i}^*),
\end{equation}
where we use the fact that the mechanism is incentive-compatible.
The payoff of the SE becomes:
\begin{equation}\label{SE_pay_as}
\Pi_0 = V - \sum_{i=1}^Nt_{i\Theta_i}(Q_i^*).
\end{equation}
Therefore, to select the optimal transfer functions, the SE must solve:
\begin{equation}\label{SE_problem_as}
\max_{t(\cdot,\cdot)}\mathbb{E}\left[u_0(\Pi_0)\right],
\end{equation}
subject to the incentive compatibility constraints of Eq.~(\ref{ic}), and the participation constraints:
\begin{equation}\label{eq:new_ic}
    \mathbb{E}_{i\theta_i}[U_i(\Pi_i(e^*_{i\Theta_i\Theta_i},\Theta_i))] \ge \bar{u}_{i\theta_i},
\end{equation}
for all $\theta_i\in\boldsymbol{\Theta}_i$, where we also assume that the incentive compatibility constrains hold.

\subsection{Parameterization of the transfer functions}
\label{sec:t_param}
Transfer functions must be practically implementable.
That is, they must be easily understood by the agent when expressed in the form of a contract.
To be easily implementable, transfer functions should be easy to convey in the form of a table.
To achieve this, we restrict our attention to functions that are made out of constants, step functions, linear functions, or combinations of these.

Despite the fact that including such functions would likely enhance the principal's payoff, we exclude transfer functions that encode penalties for poor agent performance, i.e., transfer functions that can take negative values.
First, contracts with penalties may not be implementable if the principal and the agent reside within the same organization.
Second, even when the agent is an external contractor penalties are not commonly encountered in practice.
In particular, if the SE is a sensitive government office, e.g., the department of defense, national security may dictate that the contractors should be protected from bankruptcy.
Third, we do not expect our theory to be empirically valid when penalties are included since, according to prospect theory \cite{prospect}, humans perceive losses differently.
They are risk-seeking when the reference point starts at a loss and risk-averse when the reference point starts at a gain.

To overcome these issues we restrict our attention to transfer functions that include three simple additive terms: a constant term representing a participation payment, i.e., a payment received for accepting to be part of the project; a constant payment that is activated when a requirement is met; and a linear increasing part activated after meeting the requirement.
The role of the latter two part is to incentivize the agent to meet and exceed the requirements.

We now describe this parameterization mathematically.
%Assume that agent $i=1,\dots,N$ has $M_i\ge 1$ different possible types denoted by $\theta_{ij}, j=1,\dots,M_i$.
The transfer function associated with type $k$ in $\boldsymbol{\Theta_i}$ of agent $i$ is parameterized by:
\begin{equation}
\label{basis}
\begin{array}{ccc}
t_{ik}\left(q_i\right) &=& a_{ik,0} + a_{ik,1}\hev\left(q_i - a_{ik,2}\right) \\
&&+ a_{ik,3} \left(q_i - a_{ik,2}\right)\hev\left(q_i-a_{ik,2}\right),
\end{array}
\end{equation}
where $H$ is the Heaviside function ($H(x) = 1$ if $x\ge 0$ and $0$ otherwise), and all the parameters $a_{ik,0},\dots,a_{ik,3}$ are non-negative.
In Eq.~(\ref{basis}), $a_{ik,0}$ is the participation reward, $a_{ik,1}$ is the award for exceeding the passed-down requirement, $a_{ik,2}$ is the passed-down requirement, and $a_{ik,3}$ the payoff per unit quality exceeding the passed-down requirement. 
We will call these form of transfer functions the ``requirement based plus incentive'' (RPI) transfer function. In case the $a_{ik,3}=0$, we call it the ``requirement based'' (RB) transfer function.
At this point, it is worth mentioning that the passed-down requirement $a_{ik,2}$ is not necessarily the same as the true system requirement $r_i$, see our reults in Sec. \ref{res}.
As we have shown in earlier work \cite{cesun}, the optimal passed-down requirement differs from the true system requirement. For example, the SE should ask for higher requirements for the design task with low-complexity. On the other hand, for the task with high-complexity, the SE should pass down less than the actual requirement.
For notational convenience, we denote by $\mathbf{a}_{ik}\in \mathbb{R}^{4}_+$ ($\mathbb{R} = \{x\in\mathbb{R}: x\ge 0\}$) the transfer parameters pertaining to agent $i$ of type $k\in\boldsymbol{\Theta}_i$, i.e.,
\begin{equation}\label{a_ij}
    \mathbf{a}_{ik} = \left(a_{ik,0},\dots,a_{ik,3}\right).
\end{equation}
Similarly, with $\mathbf{a}_i \in \mathbb{R}^{4M_i}_+$ we denote the transfer parameters pertaining to agent $i$ for all types, i.e.,
\begin{equation}\label{a_i}
    \mathbf{a}_i = \left(\mathbf{a}_{i1}, \dots, \mathbf{a}_{iM_i}\right),
\end{equation}
and with $\mathbf{a}\in \mathbb{R}^{4\sum_{i=1}^NM_i}_+$ all the transfer parameters collectively, i.e.,
\begin{equation}\label{a}
    \mathbf{a} = \left(\mathbf{a}_1,\dots,\mathbf{a}_N\right).
\end{equation}

\subsection{Numerical solution of the optimal contract problem}
%The maximization problem of Eq.~(\ref{eq_seopt}) subject to the constraints of Eqs.~(\ref{eq_partcons}) and~(\ref{exsub}) is analytically intractable.
%Note that, the incentive constraint in Eq.~(\ref{exsub}) is an optimization problem solved by the principal.
%Therefore, 
%The optimization problem of Eq.~(\ref{eq_seopt}),  with constraints defined by Eqs.~(\ref{eq_partcons}) and~(\ref{exsub}), is an intractable bi-level programming problem~\cite{bard}.
The optimal contract problem is a an intractable bi-level, non-linear programming problem.%~\cite{bard}.
%We solve it numerically for a finite number of agent types and within the class of implementable transfer functions, see Sec.~\ref{sec:t_param}.
In particular, the SE's problem is for the case of type-dependent contracts is to maximize the expected system-level utility over the class of implementable contracts, i.e.,
\begin{equation}\label{eq:opt}
    \underset{\mathbf{a}}{\max}\mathbb{E}\left[u_0(\Pi_0)\right],
\end{equation}
subject to
\begin{enumerate}
\item \emph{contract implementability constraints}:
\begin{equation}
a_{ik,j}\ge 0,
\label{eq:par_cons}
\end{equation}
for all $i=1,\dots,N, k=1,\dots,M_i,j=0,\dots,3$;
\item \emph{individual rationality constraints}:
\begin{equation}\label{eqn:individual-rationality}
    e^*_{ikl} = \underset{e_i\in[0,1]}{\arg\max}\mathbb{E}_{ik}[U_i(\Pi_i(e_i, l))],
\end{equation}
for all $i=1,\dots,N, k=1,\dots,M_i,l=1,\dots,M_i$;
\item \emph{participation constraints}:
\begin{equation}
    \mathbb{E}_{ik}[U_i(\Pi_i(e^*_{ikk}, k))] \ge \bar{u}_{ik},
\end{equation}
for all $i=1,\dots,N, k=1,\dots,M_i$; and
\item \emph{incentive compatibility constraints}:
\begin{equation}
    \mathbb{E}_{ik}[U_i(\Pi_i(e^*_{ikk}, k))] \ge \mathbb{E}_{ik}[U_i(\Pi_i(e^*_{ikl},l))],
\label{eq:comp_cons}
\end{equation}
for all $i=1,\dots,N$ and $k\not=l$ in $\{1,\dots,M_i\}$.
\end{enumerate}
For the case of type-independent contracts, one adds the constraint $\mathbf{a}_{ik} = \mathbf{a}_{il}$ for all $i=1,\dots,N$ and $k\not=l$ in $\{1,\dots,M_i\}$ and the incentive compatibility constraints are removed. 

%\begin{equation}
%\begin{split}
%\underset{\mathbf{a}} \max\ \ex_\omega &\left[u_0\left(\Pi_0\left(\cdot\right)\right)\right],\\
%&\text{s.t.}\ \ 
%a_{ij,k} \ge 0\ \ \text{for }j=1,\dots,M,\text{and }k=0,\cdots,3 \\ 
%&\qquad e_i^* \in \underset{e_i\in[0,1]}\argmax\ \mathbb{E}_{\rsp}\left[u_i\left(\Pi_i\left(e_i,\rsp\right)\right)\rvert \Theta_i\right],\\
%&\qquad \qquad \text{s.t.}\  \mathbb{E}_{\rsp}\left[u_i\left(\Pi_i\left(e_i^*,\rsp\right)\right)\rvert \Theta_i\right] \ge 0,\\
%&\qquad \text{for}\ i=1,\dots, N
%\end{split}
%\label{eq:opt-old}
%\end{equation}
%where,
%$$
%\mathbf{a} = \left[\mathbf{a_1},\dots,\mathbf{a_N}\right].
%$$

%\subsection{Numerical Implementation of the Bi-level Programming Problem}
A common approach to solving bi-level programming problems is to replace the internal optimization with the corresponding Karush-Kuhn-Tucker (KKT) condition. 
This approach is used when the internal problem is concave, i.e., when it has a unique maximum.
However, in our case, concavity is not guaranteed, and we resort to nested optimization.
We implement everything in Python using the Theano~\cite{theano} symbolic computation package exploit automatic differentiation.
We solve the follower problem using sequential least squares programming (SLSQP) as implemented in the scipy package. \cmmnt{\cite{sci}}
We use simulated annealing to find the global optimum point of the leader problem. We first convert the constraint problem to the unconstrained problem using the penalty method such that:
\begin{equation}
    f\left(\mathbf{a}\right)=\mathbb{E}\left[u_0(\Pi_0)\right]+\sum_{i=1}^{N_c}\min\left(g_i\left(\mathbf{a} \right), 0\right),
\label{eq:dist}
\end{equation}
where $g_i\left(\cdot\right)$'s are the constraints in Eqs. (\ref{eq:par_cons}-\ref{eq:comp_cons}). 
Maximizing the $f(\mathbf{a})$ in Eq. \ref{eq:dist}, is equivalent to finding the mode of the distribution:
\begin{equation}
    \pi_\gamma\left(\mathbf{a}\right) \propto \exp\left(\gamma f\left(\mathbf{a}\right)\right),
\end{equation}
we use Sequential Monte Carlo (SMC) \cite{smc} method to sample from this distribution by increasing $\gamma$ from $0.001$ to $50$. To perform the SMC, we use the ``pysmc'' package \cite{pysmc}.
To ensure the computational efficiency of our approach, we need to use a numerical quadrature rule to approximate the expectation over $\Xi$. 
This step is discussed in Appendix~\ref{ap:expectations}.
To guarantee the reproducibility of our results, we have published our code in an open source \href{https://github.com/ebilionis/incentives}{Github repository} (\url{https://github.com/ebilionis/incentives}) with an MIT license.

%To derive the Jacobian of the objective function (the expected utility of the principal), we calculate all the analytic derivatives with automatic differentiation using Python package, Theano~\cite{theano}.
%The Jacobian of the principal objective function which is the leading maximization problem is:
\subsection{Value Function and Risk Behavior}
\label{sec:value}
\begin{figure}[h!]
\centering
\includegraphics[scale=1]{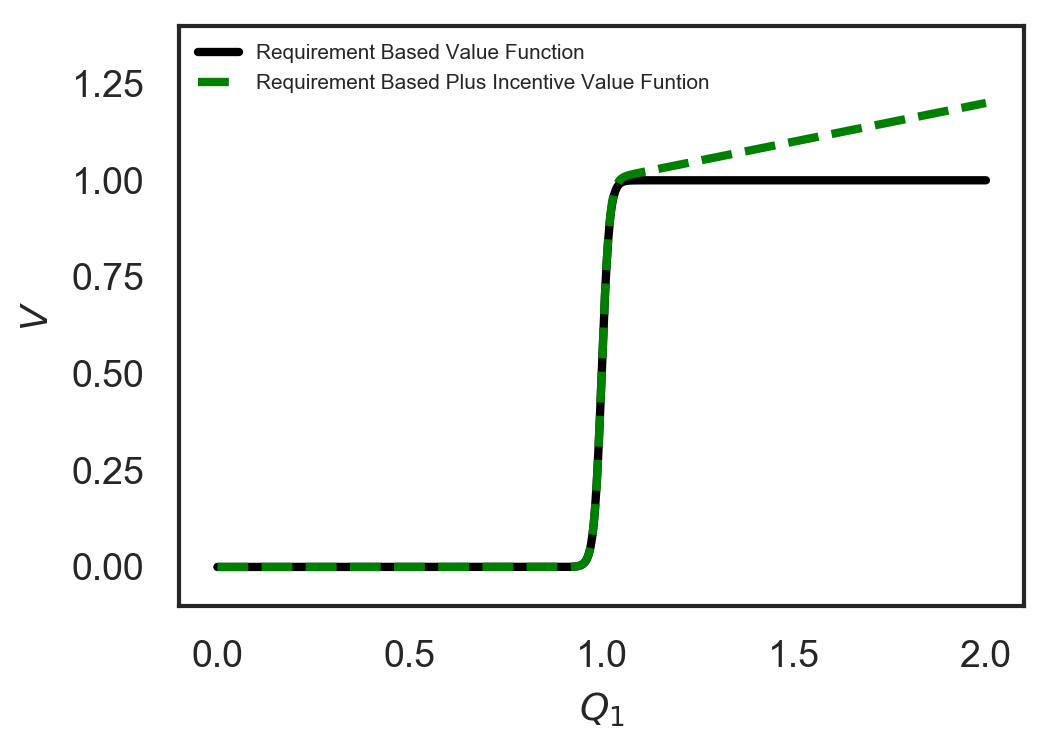}
\caption{The RB value function (black solid line) and RPI value function (green dashed line).}
\label{fig:value}
\end{figure}

We assume two types of value functions, namely, the requirement based (RB) and requirement based plus incentive (RPI). Mathematically, we define these two value functions as:
\begin{equation}
V_{\text{RB}} := v_0\prod_{i=1}^N\left\{H(Q_i^*-1)\right\}.
\end{equation}
and,
\begin{equation}
V_{\text{RPI}} := v_0\prod_{i=1}^N\left\{H(Q_i^*-1)\right\}\left[1+0.2(Q_i^*-1)\right],
\end{equation}
respectively. In Fig. \ref{fig:value}, we show these two value functions for one subsystem.
%The second value function also pays $v_0>0$ if all requirements are met, but it increases in a concave way as the quality increases.
%We define it to be:
%\begin{equation}
%V_{\text{RBI}} := v_0\prod_{i=1}^N\left\{H(Q_i^*-1)\left[1+\alpha_i\tanh(Q_i^*-1)\right]\right\}.
%\end{equation}

\begin{figure}[h]
\centering
\includegraphics[scale=1]{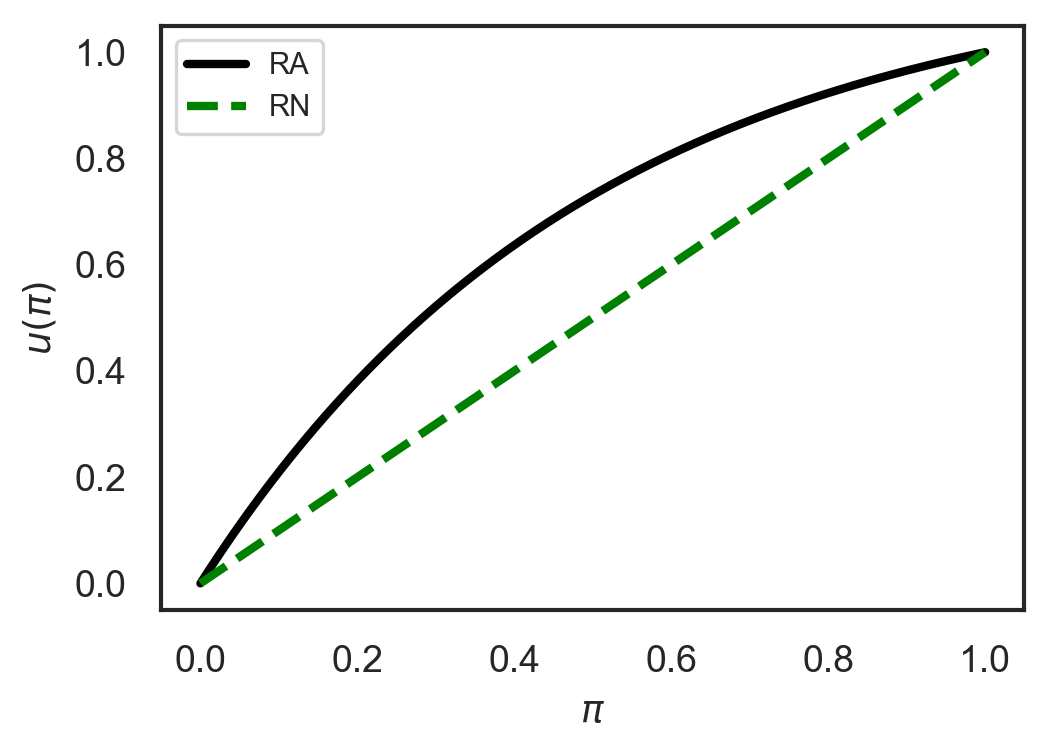}
\caption{The utility functions for risk averse (RA) (black solid line), risk neutral (RN) (green dashed line).}
\label{fig:util}
\end{figure}

We consider two different risk behaviors for individuals, risk averse (RA) and risk neutral (RN). We use the utility function in Eq. (\ref{util}), for the risk behavior of the agents and principal,
\begin{equation}
u\left(\pi\left(\cdot\right)\right) = \begin{cases} a - b e^{-c\pi\left(\cdot\right)},\ \text{for RA} \\
\pi\left(\cdot\right),\ \text{for RN},
\end{cases}
\label{util}
\end{equation}
where $c=2$ for a RA agent. The parameters $a$ and $b$ are:
$$
a=b=\frac{1}{1- e^{-c}}.
$$
We show these utility functions for the two different risk behaviors in Fig. \ref{fig:util}.

\section{NUMERICAL EXAMPLES}
\label{res}
In this section, we start by performing an exhaustive numerical investigation of the effects of task complexity, agent's cost of effort, uncertainty in the quality of the returned task, and adverse selection. 
In Sec. \ref{sec:mhrb}, we study the ``moral hazard only'' scenario with the RB transfer and value functions. In Sec. \ref{sec:mhrpi}, we study the effect of the RPI transfer and value functions. We study the ``moral hazard with adverse selection'' in Sec. \ref{sec:adcase}.

\subsection{Numerical investigation of the proposed model}
In these numerical investigations we consider a single risk neutral principal and a risk averse agent.
Each case study corresponds to a choice of task complexity ($\kappa$ in Eq. (\ref{eq:qual})), cost of effort ($c$ in Eq. (\ref{eq:cost})), and performance uncertainty ($\sigma$ in Eq. (\ref{eq:qual})).
With regards to task complexity, we select $\kappa=2.5$ for an easy task and $\kappa=1.5$ for a hard task.
For the cost of effort parameter, we associate $c=0.1$ and $c=0.4$ with the low- and high-cost agents, respectively. 
Finally, low- and high-uncertainty tasks are characterized by $\sigma=0.1$ and $\sigma=0.4$, respectively.

Note that, the parameters $\kappa_{i\theta_i}$, $c_{i\theta_i}$, and $\sigma_{i\theta_i}$ have two indices.
The first index $i$ is the agent's (subsystem's) number and the second index is the type of the agent.
We begin with a series of cases with a single agent with a known type denoted by $1$ (moral-hazard-only case studies).
In these cases, the parameters corresponding to complexity, cost and uncertainty are denoted by $\kappa_{11}$, $c_{11}$, and $\sigma_{11}$, respectively.
We end with a series of cases with a single agent but with an unknown type that can take two discrete, equally probable values $1$ and $2$ (moral-hazard-and-adverse-selection case studies).
Consequently, $\kappa_{11}$ denotes the effort coefficient of a type-1 agent $1$, $\kappa_{12}$ the same for a type-2 agent, and so on for all the other parameters.

To avoid numerical difficulties and singularities, we replace all Heaviside functions with a sigmoids, i.e., 
\begin{equation}
\hat{H}_{\alpha}(x) = \frac{1}{1 + e^{-\alpha x}},
\end{equation}
where the parameter $\alpha$ controls the slope.
We choose $\alpha=50$ for the transfer functions and $\alpha=100$ for the value function.
We consider two types of value functions, RB and RPI value functions, see Sec. \ref{sec:value}. 
For the RB value function we use the transfer function of Eq. (\ref{basis}) constrained so $a_{ik},3=0$ (RB transfer function).
In other words, the agent is paid a constant amount if they achieve the requirement and there is no payment per quality exceeding the requirement.
%The choice of this transfer function is reasonable since the value function is a Heaviside function.
For the case of RPI value function, we remove this constraint.

\subsubsection{Moral hazard with RB transfer and value functions}
\label{sec:mhrb}
\begin{figure*}[h]
\centering
	\begin{subfigure}{0.48\textwidth}
		\includegraphics[width=\textwidth]{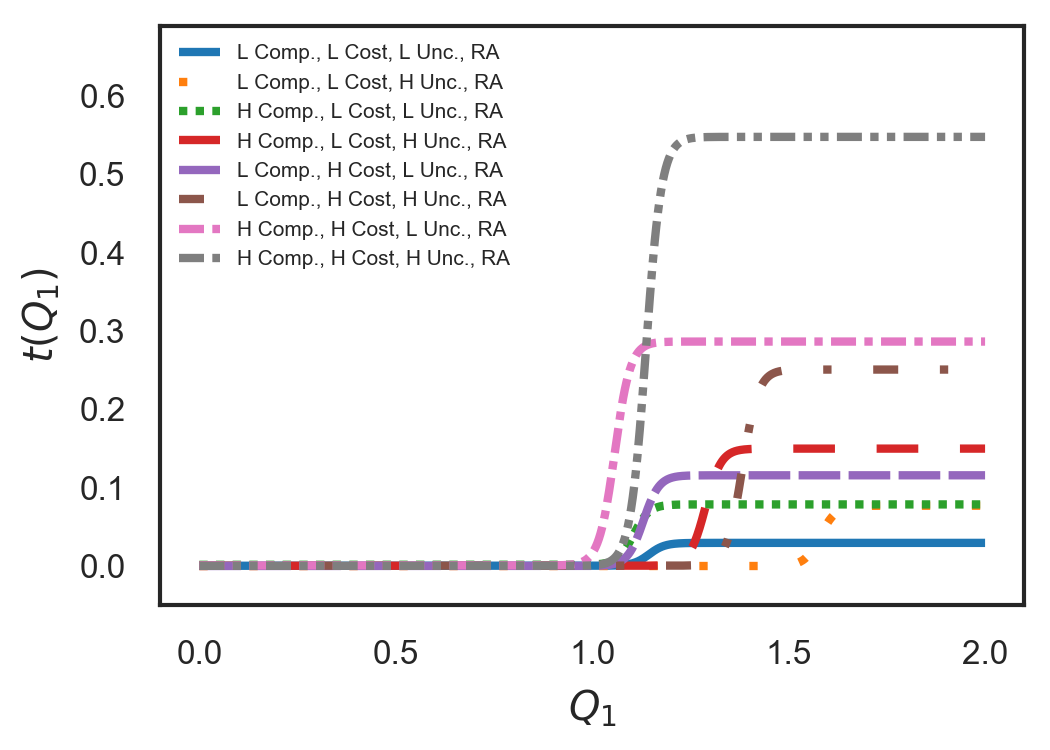}
        \caption{
        \centering
        Transfer functions using RB value function.}
        \label{fig:morals_tr}
	\end{subfigure}
	\begin{subfigure}{0.48\textwidth}
    		\includegraphics[width=\textwidth]{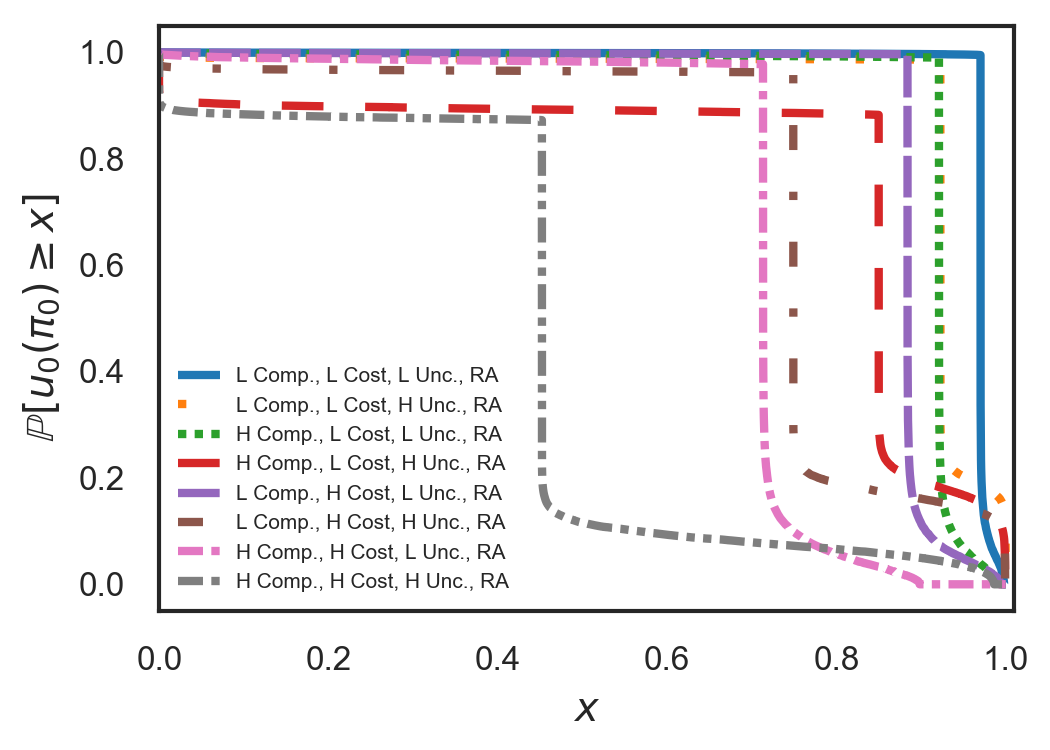}
        \caption{
        \centering
        Exceedance curve.}
        \label{fig:morals_cdf}
	\end{subfigure}
\caption{L and H stand for low and high, respectively, Comp. and Unc. stand for complexity and uncertainty, respectively. The low and high complexity denote the $\kappa_{11}=2.5$ and $\kappa_{11}=1.5$, respectively, low and high cost denote $c_{11}=0.1$ and $c_{11}=0.4$, respectively, low and high uncertainty denote $\sigma_{11}=0.1$ and $\sigma_{11}=0.4$, respectively, RA denotes the risk averse agent. (a): The RB transfer functions for several different agent types with respect to outcome of the subsystem ($Q_1$) for moral hazard scenario. (b): The exceedance for the moral hazard scenario using the RB transfer function.}
\label{fig:morals}
\end{figure*}

Consider the case of a single risk-averse agent of known type and a risk-neutral principal with an RB value function.
In Fig.~\ref{fig:morals_tr}, we show the transfer functions for several agent types covering all possible combinations of low/high complexity, low/high cost, and low/high task uncertainty.
Fig.~\ref{fig:morals_cdf} depicts the probability that the principal's expected utility exceeds a given threshold for all these combinations.
We refer to this curve as the \emph{exceedance curve}.
Finally, in tables \ref{table:moral_low} and \ref{table:moral_high}, we report the expected utility of the principal for the low and high cost agents, respectively.
We make the following observations:
\begin{enumerate}
    \item For the same level of task complexity and uncertainty, but with increasing cost of effort:
    \begin{enumerate}
        \item the optimal passed-down requirement decreases;
        \item the optimal payment for achieving the requirement increases;
        \item the principal's expected utility decreases; and
        \item the exceedance curve shifts to the left.
    \end{enumerate}
    Intuitively, as the agent's cost of effort increases, the principal must make the contract more attractive to ensure that the participation constraints are satisfied.
    As a consequence, the probability that the principal's expected utility exceeds a given threshold decreases.
    \item For the same level of task uncertainty and cost of effort, but with increasing complexity:
    \begin{enumerate}
        \item the optimal passed-down requirement decreases;
        \item the optimal payment for achieving the requirement increases;
        \item the principal's expected utility decreases; and
        \item the exceedance curve shifts to the left.
    \end{enumerate}
    Thus, we see that an increase in task complexity has the same a similar effect as an increase in the agent's cost of effort.
    As in the previous case, to make sure that the agent wants to participate, the principal has to make the contract more attractive as task complexity increases.
    \item For the same level of task complexity and cost of effort, but with increasing uncertainty:
    \begin{enumerate}
        \item the optimal passed-down requirement increases;
        \item the optimal payment for achieving the requirement increases;
        \item the principals expected utility decreases;
        \item the exceedance curve shifts towards the bottom right.
    \end{enumerate}
    This case is the most interesting.
    Here as the uncertainty of the task increases, the principal must increase the passed-down requirement to ensure that they are hedged against failure.
    At the same time, however, they must also increase the payment to ensure that the agent still has an incentive to participate.
    \item For all cases considered, the optimal passed down requirement is greater than the true requirement (which is set to one).
    Note, however, this is not universally true.
    Our study does not examine all possible combinations of cost, quality, and utility functions that could have been considered.
    Indeed, as we showed in our previous work \cite{cesun}, there are situations in which a smaller-than-the-true requirement can be optimal.
\end{enumerate}
%The results show that, for the higher cost, the optimal passed down requirement is lower than those of the lower cost with same task complexity and uncertainty. 
%Also, for the aforementioned cases, the payment for achieving the requirement is greater for the higher cost than the lower cost agents. 
%For the high complexity, the optimum passed down requirement is smaller and the award for satisfying the requirement is greater than those of the lower complexity with similar agent cost and task uncertainty. 
%Moreover, we see a clear trend that, for the higher uncertainty the optimum passed-down requirement and the payment for achieving the requirement are higher than those of the lower uncertainties while the the task complexity and agent cost are same. 

%In Fig.~\ref{fig:morals_cdf}, we show the probability of exceeding a certain amount of utility for the SE under the different scenarios for the task complexities and agent types. 
%The probability of exceeding a certain probability decreases as the task complexity or agent cost grows. 
%Moreover, the probability of exceeding a certain expected utility is smaller as the uncertainty grows. 

%In tables \ref{table:moral_low} and \ref{table:moral_high}, we show the expected utility of the principal for the low and high costs agents, respectively. As we expect, the expected utility of the principal reduces as the cost of the agents grows. Furthermore, the principal's expected utility decreases as the uncertainty of the task becomes bigger. We can also see that, the principal gains less when the the task complexity is high.

\begin{table}[h]
\centering
\caption{The expected utility of the principal for low cost agent with RB value function.}
\begin{tabular}{|c|c|c|} 
 \hline
  & Low Uncertainty & High Uncertainty \\ [0.5ex] 
 \hline
 Low Complexity & 0.97 & 0.93 \\ 
 High Complexity & 0.92 & 0.79 \\[1ex] 
 \hline
\end{tabular}
\label{table:moral_low}
\end{table}
\begin{table}[h]
\centering
\caption{The expected utility of the principal for high cost agent with RB value function.}
\begin{tabular}{|c|c|c|} 
 \hline
  & Low Uncertainty & High Uncertainty \\ [0.5ex] 
 \hline
 Low Complexity & 0.89 & 0.77 \\ 
 High Complexity & 0.72 & 0.45 \\[1ex] 
 \hline
\end{tabular}
\label{table:moral_high}
\end{table}
\subsubsection{Moral hazard with RPI transfer and value functions}
\label{sec:mhrpi}
\begin{figure*}[h]
\centering
	\begin{subfigure}{0.48\textwidth}
		\includegraphics[width=\textwidth]{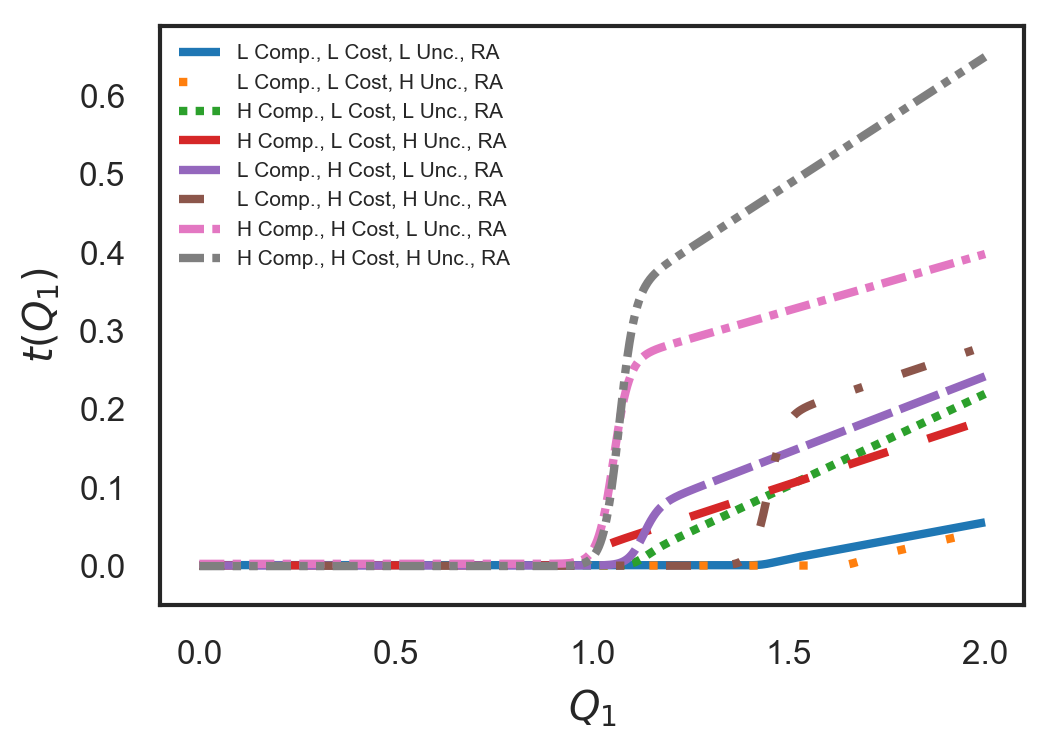}
        \caption{
        \centering
        Transfer functions using RPI value function.}
        \label{fig:morals_tr_rpi}
	\end{subfigure}
	\begin{subfigure}{0.48\textwidth}
		\includegraphics[width=\textwidth]{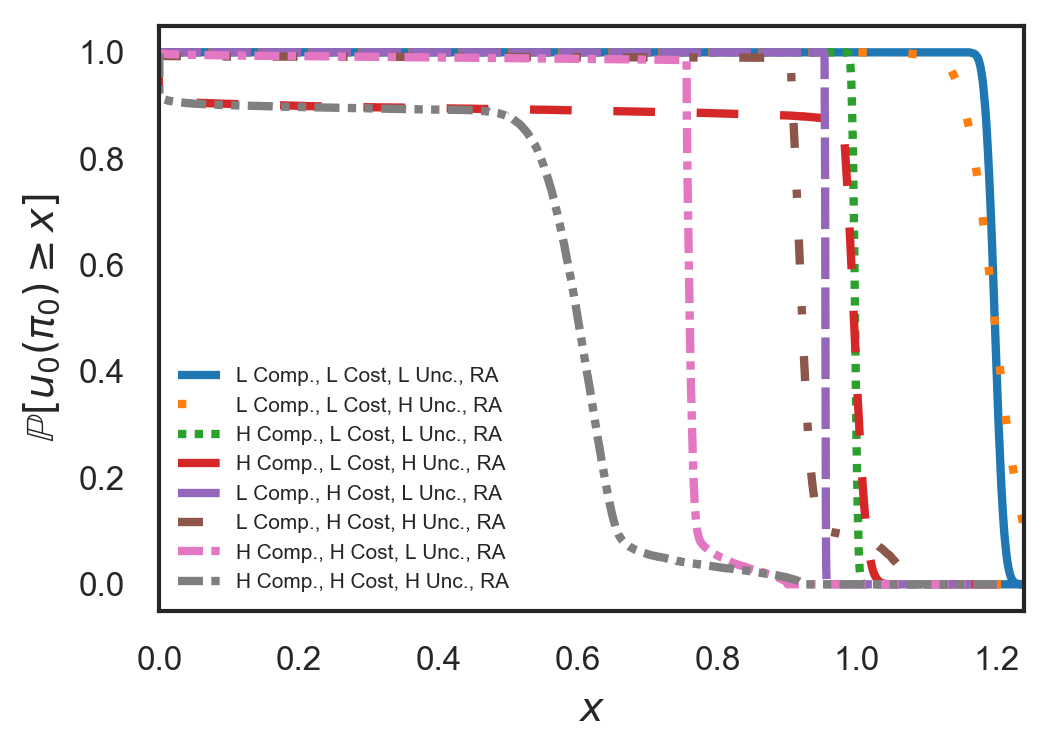}
        \caption{
        \centering
       Exceedance curve.}
        \label{fig:morals_cdf_rpi}
	\end{subfigure}
\caption{L and H stand for low and high, respectively, Comp. and Unc. stand for complexity and uncertainty, respectively. The low and high complexity denote the $\kappa_{11}=2.5$ and $\kappa_{11}=1.5$, respectively, low and high cost denote $c_{11}=0.1$ and $c_{11}=0.4$, respectively, low and high uncertainty denote $\sigma_{11}=0.1$ and $\sigma_{11}=0.4$, respectively, RA denotes the risk averse agent. (a): The RPI transfer functions for several different agent types with respect to outcome of the subsystem ($Q_1$) for moral hazard scenario. (b): The exceedance curve for the moral hazard scenario using the RPI transfer function.}
\label{fig:morals_rpi}
\end{figure*}

This case is identical to Sec.~\ref{sec:mhrb}, albeit we use the RPI value function, see Sec. \ref{sec:value}, and the RPI transfer function, see Eq. (\ref{basis}). 
Fig \ref{fig:morals_tr_rpi}, depicts the transfer functions for all combinations of agent types and task complexities.
In Fig. \ref{fig:morals_cdf_rpi}, we show the exceedance curve using the RPI value and transfer functions. 
Finally, in tables \ref{table:moral_low_RPI} and \ref{table:moral_high_RPI}, we report the expected utility of the principal using the RPI transfer and value functions for the low and high cost agents, respectively.
The results are qualitative similar to Sec.~\ref{sec:mhrb}, with the additional observations:
\begin{enumerate}
    \item For the same level of task complexity, uncertainty and agent cost, the optimal reward for achieving the requirement decreases compared to the same cases in Sec.~\ref{sec:mhrb}.
Intuitively, as the principal has the option to reward the agent based on the quality exceeding the requirement, they prefer to pay less for fulfilling the requirement. 
Instead, the principal incentivizes the agent to improve the quality beyond the optimal passed-down requirement.
\item The slope of the transfer function beyond the passed-down requirement is almost identical to the slope of the value function.
\end{enumerate}

\begin{table}[h!]
\centering
\caption{The expected utility of the principal for low cost agent with RPI value function.}
\begin{tabular}{|c|c|c|} 
 \hline
  & Low Uncertainty & High Uncertainty \\ [0.5ex] 
 \hline
 Low Complexity & 1.2 & 1.2 \\ 
 High Complexity & 1.0 & 0.89 \\[1ex] 
 \hline
\end{tabular}
\label{table:moral_low_RPI}
\end{table}
\begin{table}[h!]
\centering
\caption{The expected utility of the principal for high cost agent with RPI value function.}
\begin{tabular}{|c|c|c|} 
 \hline
  & Low Uncertainty & High Uncertainty \\ [0.5ex] 
 \hline
 Low Complexity & 0.95 & 0.93 \\ 
 High Complexity & 0.76 & 0.56 \\[1ex] 
 \hline
\end{tabular}
\label{table:moral_high_RPI}
\end{table}
In table \ref{summary}, we summarize our observations for the results in Sec. ~\ref{sec:mhrb} and \ref{sec:mhrpi}. 
In this table, we show how the passed-down requirement and payment change when we fix two parameters of the model (we denote it by ``fix'' in the table) and vary the third parameter.
We denote increase by $\uparrow$ and decrease by $\downarrow$.
\begin{table}[h!]
\centering
\caption{Summary of the observations.}
\begin{tabular}{|c|c|c|c|c|} 
 \hline
  complexity & agent cost & uncertainty & requirement & payment \\ [0.5ex] 
 \hline
 $\uparrow$ & fix & fix & $\downarrow$ & $\uparrow$ \\ 
 fix & $\uparrow$ & fix & $\downarrow$ & $\uparrow$ \\
 fix & fix & $\uparrow$ & $\uparrow$ & $\uparrow$ \\ 
 [1ex] 
 \hline
\end{tabular}
\label{summary}
\end{table}
\subsubsection{Moral hazard with adverse selection}
\label{sec:adcase}
Consider the case of a single risk-averse agent of unknown type which takes two possible values, and a risk-neutral principal with a RB value function.
We consider two possibilities for the unknown type:
\begin{enumerate}
    \item \emph{Unknown cost of effort.}
    Here, we set $\kappa_{11}=\kappa_{12}=1.5$ ($p(\kappa_{11}=1.5) = 1$), $\sigma_{11}=\sigma_{12}=0.1$ ($p(\sigma_{11}=0.1)$), and $p(c_{11}=0.1)=0.5$ and $p(c_{12}=0.4)=0.5$
    \item \emph{Unknown task complexity.} For the unknown quality we assume that $p(\kappa_{11}=2.5)=0.5$ and $p(\kappa_{12}=1.5)=0.5$, $\sigma_{11}=\sigma_{12}=0.4$ ($p(\sigma_{11}=0.4)=1$), and $c_{11}=c_{12}=0.4$ ($p(c_{11}=0.4)=1$). 
\end{enumerate}
In this scenario, we maximize the expected utility of the principal subject to constraints in Eqs. (\ref{eq:par_cons}-\ref{eq:comp_cons}). 
The incentive compatibility constraint, Eq. (\ref{eq:comp_cons}), guarantees that the agent will choose the contract that is suitable for their true type. 
In other words, as there are two agent types' possibilities, the principal must offer two contracts, see Fig. \ref{fig:timing2}.
These two contracts must be designed in a way that there is no benefit for the agent to deviate from their true type, i.e., the contracts enforce the agent to be truth telling. 

Solving the constraint optimization problem yields:
$$
\mathbf{a}_{11}=\mathbf{a}_{12}=(0, 0.29, 1.06),
$$
i.e., the two contracts collapse into one.
Note that the resulting contract is the same as the pure moral hazard case, Sec. \ref{sec:mhrb}, for an agent with type $\kappa_{11}=1.5$, $\sigma_{11}=0.1$, and $c_{11}=0.4$.
%$$
%\mathbf{a}_{11}=(0, 0.2858, 1.0570).
%$$
In other words, the principal must behave as if there was only a high-cost agent.
That is, there are no contacts that can differentiate between a low- and a high-cost agent in this case.

A similar outcome occurs for unknown task complexity.
The solution of the constraint optimization problem for this scenario is:
$$
\mathbf{a}_{11}=\mathbf{a}_{12}=(0, 0.08, 1.11),
$$
which is the same as the optimum contract that is offered for the pure moral hazard case, Sec. \ref{sec:mhrb}, for an agent with type $\kappa_{11}=1.5$, $\sigma_{11}=0.4$, and $c_{11}=0.4$.
Therefore, in this case the principal must behave as if there the task is of high complexity. 

Note that in both cases above, the collapse of the two contracts to one contract is not a generalizable property of our model.
In particular, it may not happen if more flexible transfer functions are allowed, e.g., ones that allow performance penalties.

\begin{figure}[h]
\centering
\includegraphics[scale=1.0]{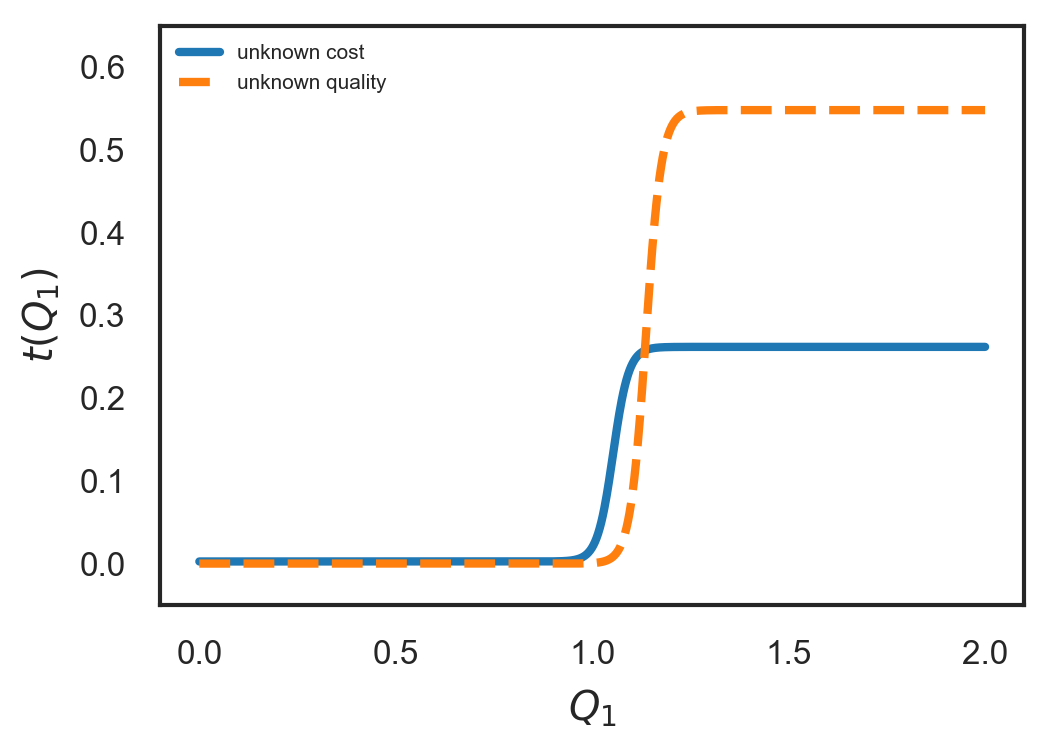}
\caption{The transfer function for the adverse selection scenarios with unknown cost (solid line) and unknown quality (dashed line), the agent is risk averse. For unknown cost: $\kappa_{11}=\kappa_{12}=1.5$ with probability 1, $\sigma_{11}=\sigma_{12}=0.1$ with probability 1, and $c_{11}=0.1$ with probability 0.5 and $c_{12}=0.4$ with probability 0.5. For unknown quality: $\kappa_{11}=2.5$ with probability 0.5 and $\kappa_{12}=1.5$ with probability 0.5, $\sigma_{11}=\sigma_{12}=0.4$ with probability 1, and $c_{11}=c_{12}=0.4$ with probability 1.}
\label{fig:ads}
\end{figure}

In Fig. \ref{fig:ads}, we show the transfer functions for the adverse selection scenarios with unknown cost and unknown quality. 
In tables \ref{table:ad_cost} and \ref{table:ad_qual}, we show the expected utility of two types of agents and the principal using the optimum contract for unknown cost and unknown quality, respectively.
To sum up:
\begin{enumerate}
    \item The unknown cost:
    \begin{enumerate}
        \item the optimum transfer function for this problem is as same as that the principal would have offered for a single-type high-cost agent with $c_{11}=c_{12}=0.4$ (moral hazard scenario with no adverse selection);
        %This may not happen for other probabilities; 
        \item the expected utility of the low cost agent (efficient agent) is greater than that of the high cost agent.
    \end{enumerate}
        In this case, the low-cost agent benefits because of information asymmetry.
        In other words, the principal must pay an information rent to the low-cost agent to reveal their type.
        \item The unknown task complexity:
    \begin{enumerate}
        \item the optimum contract in this case is the contract that the principal would have offered for the single-type high-complexity task with $\kappa_{11}=\kappa_{12}=1.5$;
        %This may not happen for other probabilities for the task complexity;
        \item the expected utility of an agent dealing with a low-complexity task is greater than that of an agent dealing with a high-complexity task.
    \end{enumerate}
    Again, due to the information asymmetry, the agent benefits if the task complexity is low.
    The principal must pay an information rent to reveal the task complexity.
\end{enumerate}

\begin{table}[h]
\centering
\caption{The expected utility of the agent with unknown cost for two different contracts.}
\begin{tabular}{|c|c|c|} 
 \hline
  & $ \mathbb{E}[u_1(\cdot)]$ &$ \mathbb{E}[u_0(\cdot)] $\\ [0.5ex] 
 \hline
 Low Cost Agent (Type 1)  & 0.39 & 0.72 \\ 
 High Cost Agent (Type 2)  & 0 & 0.72 \\[1ex] 
 \hline
\end{tabular}
\label{table:ad_cost}
\end{table}

\begin{table}[h]
\centering
\caption{The expected utility of the agent with unknown quality for two different contracts.}
\begin{tabular}{|c|c|c|} 
 \hline
  & $\mathbb{E}[u_1(\cdot)]$ & $\mathbb{E}[u_0(\cdot)]$ \\ [0.5ex] 
 \hline
 Low Complexity (Type 1)  & 0.52 & 0.45 \\ 
 High Complexity (Type 2) & 0 & 0.45 \\[1ex] 
 \hline
\end{tabular}
\label{table:ad_qual}
\end{table}
\subsection{Satellite Design}
\label{sec:sat}
In this section we apply our method on a \emph{simplified} satellite design. 
Typically a satellite consists of seven different subsystems ~\cite{wertz_space_2011}, namely, electrical power subsystem, propulsion, attitude determination and control, on-board processing, telemetry, tracking and command, structures and thermal subsystems. 
We focus our attention on the propulsion subsystem ($N=1$).
To simplify the analysis, we assume that the design of these subsystems is assigned to a sSE in a one-shot fashion.
Note that, the actual systems engineering process of the satellite design is an iterative process and the information and results are exchanged back and forth in each iteration.
Our model is a crude approximation of reality. 
The goal of the SE is to optimally incentivize the sSE to produce subsystem designs that meet the mission's requirements. 
Furthermore, we assume that the propulsion subsystem is decoupled from the other subsystems, i.e., there is no interactions between them, and that the SE knows the types of each sSE and therefore, there is no information asymmetry.

To extract the parameters of the model, i.e., $a_{11}, \sigma_{11}, c_{11}$, we will use available historical data. To this end, let $I_1$ be the cumulative, sector-wide investment on the propulsion subsystem and $G_1$ be the delivered specific impulse of solid propellants ($I_{\text{sp}}$). The specific impulse is defined as the ratio of thrust to weight flow rate of the propellant and is a measure of energy content of the propellants~\cite{wertz_space_2011}.

Historical data, say $\mathcal{D}_1 = \left\{\left(I_{1,i}, G_{1,i}\right)\right\}_{i=1}^{S}$, of these quantities are readily available for many technologies.
Of course, cumulative investment and best performance increase with time, i.e., $I_{1,i} \le I_{1,i+1}$ and $G_{1,i} \le G_{1,i+1}$.
We model the relationship between $G_1$ and $I_1$ as:
\begin{equation}
\label{G_of_I}
G_{1} = G_{1,S} + A_{1}(I_{1}-I_{1,S})+ \Sigma_{1}\Xi_1, 
\end{equation}
where  $G_{1,S}$ and $I_{1,S}$ are the current states of these variables, $\Xi_1\sim\mathcal{N}(0,1)$, and $A_1$ and $\Sigma_1$ are parameters to be estimated from the all available data, $\mathcal{D}_1$.
We use a maximum likelihood estimator for $A_1$ and $\Sigma_1$.
This is equivalent to a least squares estimate for $A_i$:
\begin{equation}
\hat{A}_1 = \arg\min_{A_1}\sum_{i=1}^{S}\left[G_{1,S}+A_1(I_{1,i}-I_{1,S}) - G_{1,i}\right]^2,
\end{equation}
and to setting $\Sigma_1$ equal to the mean residual square error:
\begin{equation}
\hat{\Sigma}_1 = \frac{1}{S}\sum_{i=1}^{S}\left[G_{1,S}+\hat{A_1}(I_{1,i}-I_{1,S}) - G_{1,i}\right]^2.
\end{equation}

Now, let $G_1^r$ be the required quality for the propulsion subsystem in physical units.
The scaled quality of a subsystem $Q_1$, can be defined as:
\begin{equation}
\label{q_def}
Q_{1} = \frac{G_1 - G_{1,S}}{G_1^r - G_{1,S}}, 
\end{equation}
with this definition, we get $Q_1=0$ for the state-of-the-art, and $Q_1=1$ for the requirement.
Substituting Eq.~(\ref{G_of_I}) in Eq.~(\ref{q_def}) and using the maximum likelihood estimates for $A_1$ and $\Sigma_1$, we obtain:
\begin{equation}
\label{q_intermediate}
Q_{1} = \frac{\hat{A}_{1}}{G_1^r - G_{1,S}}(I_{1}-I_{1,S})+ \frac{\hat{\Sigma}_{1}}{G_1^r - G_{1,S}}\Xi_1.
\end{equation}
From this equation, we can identify the uncertainty $\sigma_{11}$ in the quality function as:
\begin{equation}
\sigma_{11} = \frac{\hat{\Sigma}_{1}}{G_1^r - G_{1,S}}.
\label{eq:real_sigma}
\end{equation}

Finally, we need to define effort.
Let $T_1$ represents the time for which the propulsion engineer is to be hired. The cost of the agent per unit time is $C_1$.
$T_1$ is just the duration of the systems engineering process we consider.
The value $C_1T_1$ can be read from the balance sheets of publicly traded firms related to the technology.
We can associate the effort variable $e_1$ with the additional investment required to buy the time of one engineer:
\begin{equation}
\label{e_def}
e_{1} = \frac{I_{1}-I_{1,S}}{C_1{T_{1}}},
\end{equation}
that is, $e_1 = 1$ corresponds to the effort of one engineer for time $T_1$. Let us assume there are $Z$ engineers work on the subsystem. 
Comparing this equation, Eq.~(\ref{q_intermediate}),~and~Eq.~(\ref{eq:qual}), we get that the $\kappa_{11}$ coefficient is given by:
\begin{equation}
\kappa_{11} = \frac{Z C_1T_1 \hat{A}_1}{G_1^r-G_{1,S}}.
\label{eq:real_kappa}
\end{equation}
To complete the picture, we need to talk about the value $V_0$ (in USD) of the system if the requirements are met.
We can use this value to normalize all dollar quantities.
That is, we set:
\begin{equation}
    v_0 = 1,
\end{equation}
and for the cost per square effort of the agent we set:
\begin{equation}
    c_{11} = \frac{Z C_1T_1}{V_0}.
\label{eq:real_c}
\end{equation}

%Typically a satellite consists of seven different subsystems ~\cite{wertz_space_2011}, namely, electrical power subsystem (EPS), propulsion, attitude determination and control (ADC), on-board processing, telemetry, tracking and command (TT\&C), structures and thermal subsystems. 
%We will focus our attention on two subsystems ($N=2$): EPS and propulsion.
%To simplify the analysis, we assume that the design of these subsystems will be assigned to two sSE's in a one-shot fashion. Note that, the systems engineering process of the satellite design is an iterative process and the information and results are exchanged back and forth in each iteration.

\begin{figure}[h!]
\centering
\includegraphics[scale=1.0]{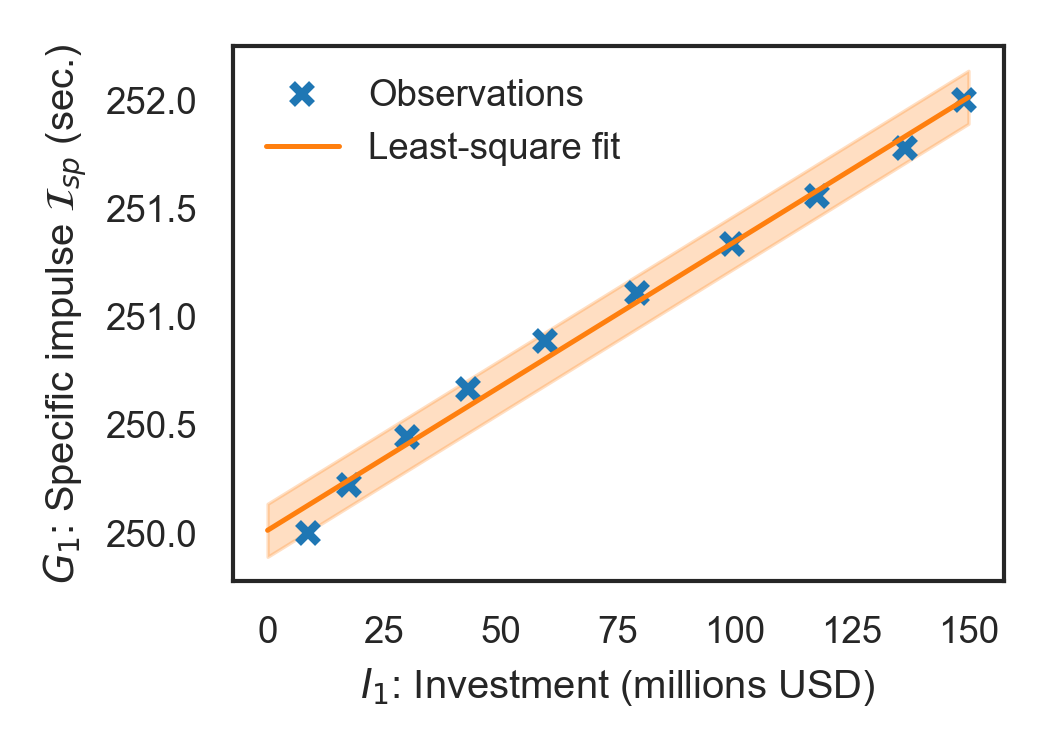}
 \caption{
 Satellite case study (propulsion subsystem): Historical data (1979--1988) of specific impulse of solid mono-propellants vs cumulative investment per firm.
The solid line and the shaded area correspond to the maximum likelihood fit of a linear regression model and the corresponding $95\%$ prediction intervals, respectively. 
}
\label{propdata3}
\end{figure}

Finally, we use some real data to fix some of the parameters.
Trends in delivered $I_{\text{sp}}$ ($G_1$ (sec.)) and investments by NASA ($I_1$ (millions USD)) in chemical propulsion technology with time are obtained from~\cite{noauthor_solid_nodate} and~\cite{noauthor_nasa_nodate}, respectively. The state-of-the-art solid propellant technology corresponds to a $G_{1,S}$ value of 252~sec. and $I_{1,S}$ value of 149.1 million USD. 
The maximum likelihood fit of the parameters results in a regression coefficient of $\hat{A}_1=0.0133$ sec. per million USD, and standard deviation $\hat{\Sigma}_1 = 0.12$ sec. The corresponding data and the maximum likelihood fit are illustrated in Fig.~\ref{propdata3}.
The value of $C_1$ is the median salary (per time) of a propulsion engineer which is approximately 120,000 USD / year, according to the data obtained from~\cite{prop_median_sal}. 
For simplicity, also assume that $T_1 = 1$ year.
Moreover, we assume that there are 200 engineers work on the subsystem, $Z=200$.
%We will examine all combinations of a low/high requirement and a low/high system value. We summarize 4 parameters for case studies in table Let us consider the case studies in table \label{sat_params}.
We will examine two case studies which is summarized in table \ref{tab:sat_params}.

\begin{figure}[h!]
\centering
\includegraphics[scale=1.0]{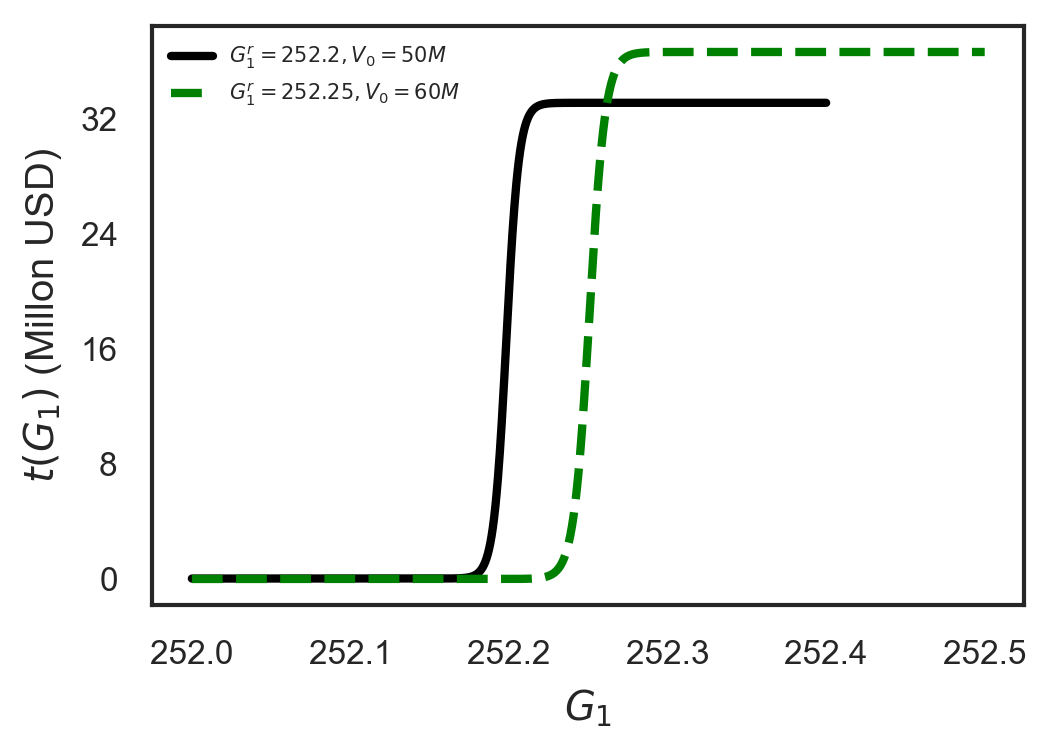}
 \caption{
 The contracts for two case studies in satellite design. 
}
\label{fig:sat_contract}
\end{figure}

\begin{table}[h!]
\centering
\caption{The model parameters for two case studies.}
\begin{tabular}{|c|c|c|c|c|} 
 \hline
  $G_1^r$ & $V_0$ & $\kappa_{11}$ & $\sigma_{11}$ & $c_{11} $\\ [0.5ex] 
 \hline
 252.2\ s & 50,000,000\ USD & 1.6 & 0.6 & 0.5\\ 
 252.25\ s & 60,000,000\ USD  & 1.28 & 0.48 & 0.4 \\[1ex] 
 \hline
\end{tabular}
\label{tab:sat_params}
\end{table}
Using RB value function, we depict the contracts for these two scenarios in Fig. \ref{fig:sat_contract}.

\section{CONCLUSIONS}
\label{sec:concl}
We developed a game-theoretic model for a one-shot shallow SEP. 
We posed and solved the problem of identifying the contract (transfer function) that maximizes the principal's expected utility. 
Our results show that, the optimum passed-down requirement is different from the real system requirement. 
For the same level of task complexity and uncertainty, as the agent cost of effort increases, the passed-down requirement decreases and the award to achieving the requirement increases. 
In this way, the principal makes the contract more attractive to the high-cost agent and ensures that the participation constraint is satisfied.
Similarly, for the same level of task uncertainty and cost of effort, increasing task complexity results in lower passed-down requirement and larger award for achieving the requirement. 
For the same level of task complexity and cost of effort, as the uncertainty increases both the passed-down requirement and the award for achieving the requirement increase. 
This is because the principal wants to make sure that the system requirements are achieved. 
Moreover, by increasing the task complexity, the task uncertainty, or the cost of effort, the principal earns less and the exceedance curve is shifted to the left.
Using the RPI contracts, the principal pays smaller amount for achieving the requirement but, instead, they pay for per quality exceeding the requirement. 

For the adverse selection scenario with RB value function, we observe that when the principal is maximally uncertain about the cost of the agent, the optimum contracts are equivalent to the contract designed for the high cost agent in the single-type case with no adverse selection. 
The low-cost agent earns more expected utility than the high-cost agent. 
This is the information rent that the principal must pay to reveal the agents' types.
Similarly, if the principal is maximally uncertain about the task complexity, the two optimum contracts for the unknown quality are equivalent to the contract that is offered to the high-complexity task where there is no adverse selection. 
Note that, the equivalence of the contracts in adverse selection scenario with the contract that is offered in absence of adverse selection is not universal.
If the class of possible contracts is enlarged, e.g., to allow penalties, there may be a set of two contracts that differentiate types.

There are still many remaining questions in modeling SEPs using a game-theoretic approach.
First, there is a need to study the hierarchical nature of SEPs with potentially coupled subsystems.
Second, true SEPs are dynamic in nature with many iterations corresponding to exchange of information between the various agents.
These are the topics of ongoing research towards a theoretical foundation of systems engineering design that accounts for human behavior.

\appendices
\section{Numerical estimation of the required expectations}
\label{ap:expectations}
For the numerical implementation of the suggested model, we need to be able to carry out expectations of the form of Eq.~\ref{eq:exp_ik} a.k.a. Eq. \ref{eq:exp} and Eq. \ref{eq:prob}.
Since, we have at most two possible types in our case studies, the summation over the possible types is trivial.
Focusing on expectations over $\Xi$, we evaluate them using a sparse grid quadrature rule \cite{sparse_grid}.
In particular, any expectation of the form $\mathbb{E}[g(\Xi)]$ is approximated by:
\begin{equation}
    \mathbb{E}[g(\Xi)] \approx \sum_{s=1}^{N_s}w^{(s)}g\left(\xi^{(s)}\right),
\end{equation}
where $w^{(s)}$ and $\xi^{(s)}$ are the $N_s = 127$ quadrature points of the level $6$ sparse grid quadrature constructed by the Gauss-Hermite 1D quadrature rule.

\section*{Acknowledgment}
This material is based upon work supported by the National Science Foundation under Grant No. 1728165.

\ifCLASSOPTIONcaptionsoff
  \newpage
\fi



%\bibliographystyle{IEEEtran}
\bibliographystyle{IEEEtran}
\bibliography{references}


%\end{thebibliography}


%\newpage


%\vfill


\end{document}